# Introducing Gaussian covariance graph models in genome-wide prediction

Carlos Alberto Martínez[*,1], Kshitij Khare[+,2], Syed Rahman[+,3], Mauricio A. Elzo[*,4]

[*]Department of Animal Sciences
[+]Department of Statistics
University of Florida, Gainesville, FL, USA

## Abstract

Several statistical models used in genome-wide prediction assume independence of marker allele substitution effects, but it is known that these effects might be correlated. In statistics, graphical models have been identified as a useful tool for covariance estimation in high dimensional problems and it is an area that has recently experienced a great expansion. In Gaussian covariance graph models (GCovGM), the joint distribution of a set of random variables is assumed to be Gaussian and the pattern of zeros of the covariance matrix is encoded in terms of an undirected graph $G$. In this study, methods adapting the theory of GCovGM to genome-wide prediction were developed (Bayes GCov, Bayes GCov-KR and Bayes GCov-H). In simulated and real datasets, improvements in correlation between phenotypes and predicted breeding values and accuracies of predicted breeding values were found. Our models account for correlation of marker effects and permit to accommodate general structures as opposed to models proposed in previous studies which consider spatial correlation only. In addition, they allow incorporation of biological information in the prediction process through its use when constructing graph $G$, and their extension to the multiallelic loci case is straightforward.

## Introduction

Most of the Bayesian and classical models used in genome-wide prediction (Meuwissen et al., 2001) assume that marker allele substitution effects follow independent Gaussian distributions which induces a diagonal covariance matrix; however, some biological phenomena point to non-independent effects. On one hand, the existence of linkage disequilibrium (LD) may create a spatial correlation of marker effects (Gianola et al., 2003; Yang and Tempelman, 2012). On the other hand, the complex interactions between regions of the genome and interactions of gene products in the metabolism also suggest that the assumption of independent effects may not be tenable. Thus, accounting for correlated marker allele substitution effects may increase the predictive performance of statistical models used in genome-wide prediction. Although it has been known that marker effects might be correlated, the problem of accounting for such a correlation has not been widely studied. So far, there have been few studies investigating this interesting problem. Gianola et al., (2003) described a series of frequentist and Bayesian models accounting for within chromosome correlated marker effects. Yang and Tempelman (2012) proposed a Bayesian antedependence model considering a nonstationary correlation structure of SNP effects. The two studies only considered correlations among nearby markers.

---

[1] carlosmn@ufl.edu
[2] kdkhare@ufl.edu
[3] Shr264@ufl.edu
[4] maelzo@ufl.edu

Covariance estimation is recognized as a challenging problem in statistics (Stein, 1975), especially in high dimensional problems under the "big $p$ small $n$" condition where the sample covariance matrix is not of full rank (Rajaratnam et al., 2008). As a consequence, high dimensional covariance estimation using graphical models is a contemporary topic in statistics and machine learning. Regularization methods imposing sparsity on estimators through structural zeros in the covariance or inverse covariance matrix have gain attention during recent years, (Carvalho et al., 2007; Letac and Massan, 2007; Rajaratnam et al., 2008). In these models, the pattern of zeros of the covariance (covariance graph models) or precision matrix (concentration graph models) is defined by means of an undirected graph $G$. The nodes of this graph represent the underlying random variables, and when the joint distribution of these variables is multivariate Gaussian, pairs of nodes not sharing an edge in $G$ are either, marginally independent (Gaussian covariance graph models) or conditionally (given all other variables) independent (Gaussian concentration graph models). This paper focuses on Gaussian covariance graph models (GCovGM). In statistics, the usefulness of these models in the analysis of high dimensional data exhibiting dependencies is well known (Carvalho et al., 2007; Rajaratnam et al., 2008); consequently, given the need for flexible statistical methods to account for correlated marker effects in genome-wide prediction, the introduction of GCovGM in this area seems promising.

To our knowledge, this is the first study adapting the theory of GCovGM to account for correlated SNP allele substitution effects in genome-wide prediction. The theory of GCovGM has been developed to estimate the covariance matrix of an observable $p$-dimensional random vector using $N$ iid observations. In contrast, in genome-wide prediction, the problem involves predicting marker effects, estimating residual variance(s), and estimating the covariance matrix of an unobservable random vector (SNP effects) using one $n$-dimensional vector with phenotypic information along with genomic information. Thus, the objective of this study was to develop methods that adapt the theory of GCovGM to genome-wide prediction in order to account for correlated marker allele substitution effects.

## Materials and methods

This section is split into the following subsections. Firstly, due to the fact that GCovGM theory is not widely known in the realm of quantitative genetics, a brief introduction and details on the challenge encountered when adapting it to genome-wide prediction are presented. Then, statistical methods adapting GCovGM to genome-wide prediction are described along with some approaches to build the graph $G$. Finally, datasets used to implement our methods are described.

**Gaussian Covariance Graph Models**

Here, the case of a known graph $G$ is considered. By known $G$ it is meant that the pattern of zeros in the covariance matrix is actually known or that $G$ is defined on the basis of domain-specific knowledge. Some basic concepts in graph theory are provided in supporting information (Appendix A); the reader not familiar with this topic is encouraged to read it before reading the rest of the paper. Hereinafter, the operator $|\cdot|$ represents the determinant when the argument is a matrix and cardinality when the argument is a set. Let $Y_1, Y_2, \ldots, Y_N$ be a set of vectors in $\mathbb{R}^p$ identically and independently distributed $MVN(0, \Sigma)$, the target is to estimate $\Sigma$. The graph $G$ determines the null entries of $\Sigma$ as explained before. Formally, the parameter space is defined as follows. Let $G = (V, E)$ be an undirected graph with vertex set $V$ and edge set $E$, then $\Sigma$ lies in the cone $\mathbb{P}_G = \{A: A \in \mathbb{P}^+ \text{ and } A_{ij} = 0 \text{ whenever } (i,j) \notin E\}$, where $\mathbb{P}^+$ is the space of positive definite matrices. Maximum likelihood estimation is possible only when $N > p$ and because of the constraints that it imposes when adapting GCovGM in genome-wide prediction (see supplementary material, Appendix B) this paper focuses on Bayesian approaches only.

*Bayesian estimation*

For natural exponential families (as in concentration graph models) a class of conjugate priors corresponding to the Diaconis-Ylvisaker prior (Diaconis and Yilvisaker, 1979) is frequently used. However, covariance graph models correspond to curved exponential families instead of natural exponential families. It is easily checked because $L(\Sigma) \propto \exp\left(-\frac{N}{2}tr(\Sigma^{-1}S) - \frac{N}{2}log|\Sigma|\right), \Sigma \in \mathbb{P}_G$, where $S$ is the sample covariance matrix, notice that $L(\Sigma)$ does not have the form of a natural exponential family. Silva and Ghahramani (2009) introduced the family of conjugate priors known as inverse $G$-Wishart ($IGW(U, \delta)$) whose pdf has the following form: $\pi_{U,\delta}(\Sigma) \propto \exp\left(-\frac{tr(\Sigma^{-1}U)}{2} - \frac{\delta}{2}log|\Sigma|\right), \Sigma \in \mathbb{P}_G$. Let $Y \coloneqq (Y_1, Y_2, \ldots, Y_N)$. Under this prior: $\pi_{U,\delta}(\Sigma|Y) \propto L(\Sigma)\pi_{U,\delta}(\Sigma) \propto \exp\left(-\frac{1}{2}tr(\Sigma^{-1}(U + NS)) - \frac{N+\delta}{2}log|\Sigma|\right), \Sigma \in \mathbb{P}_G$. This corresponds to a $IGW(\widetilde{U}, \widetilde{\delta})$ distribution, $\widetilde{U} \coloneqq U + NS, \widetilde{\delta} \coloneqq N + \delta$. An important issue that has to be considered now is for which values of $U$ and $\delta$, $\pi_{U,\delta}(\cdot)$ is a valid density. To find sufficient conditions the modified Cholesky decomposition of $\Sigma$, $\Sigma = LDL'$, where $L$ is a lower triangular matrix with diagonal entries equal to one and $D$ is a strictly positive diagonal matrix, is used. Then, we have the following bijection $\{\Sigma_{ij}\}_{i \geq j, (i,j) \in E} \to \left(\{L_{ij}\}_{i > j, (i,j) \in E}, D\right)$, and this transformation induces the density $\pi_{U,\delta}(L, D) \propto \exp\left(-tr(D^{-1}L^{-1}U(L')^{-1}) - \frac{1}{2}\sum_{j=1}^{p}(\delta + 2n_j)logD_{jj}\right)$ where $n_j = |\{i: i > j, (i,j) \in E\}| \forall j = 1, 2, \ldots, p - 1$. From the mathematical point of view, the problem is to find sufficient conditions for the following integral to be finite:

$$\int_{\mathbb{R}^{|E|}} \int_{\mathbb{R}^p_+} \pi_{U,\delta}(L, D) dD dL,$$

after some manipulations, it can be shown that these conditions are the following (Khare and Rajaratnam, 2011). 1) $U \in \mathbb{P}^+$, 2) $\delta - 2n_j > v_j + 2 \forall j = 1, 2, \ldots, p$, where $v_j = |\{i < j: (i,j) \in E\}|$. For covariance graph models, there is a block Gibbs sampler algorithm to draw samples from the posterior. This sampler is based on partitioning the covariance matrix as: $\Sigma = \begin{bmatrix} \Sigma_{11} & \Sigma'_{.1} \\ \Sigma_{.1} & \Sigma_{-1,-1} \end{bmatrix}$ and it uses the following result. Let $\boldsymbol{\beta}_1 \coloneqq (\Sigma_{1j})_{(1,j) \in E}$, i.e., a vector containing the unconstrained (non-null) covariance parameters for variable 1, $\gamma_1 = \Sigma_{11} - \Sigma'_{.1}\Sigma^{-1}_{-1,-1}\Sigma_{.1}$, and $Q_1$ = a matrix of zeros and ones such that: $\Sigma_{.1} = Q_1\beta_1$, then: $\beta_1|\gamma_1,\Sigma_{-1,-1} \sim MVN(A^{-1}Q'_1\Sigma^{-1}_{-1,-1}U_{.1}, \gamma_1 A^{-1})$ and $\gamma_1|Q_1,\beta_1,\Sigma_{-1,-1} \sim IG\left(\frac{\delta}{2} - 1, \frac{U_{11} - 2U'_{.1}\Sigma^{-1}_{-1,-1}Q_1\beta_1 + \beta'_1 A\beta_1}{2}\right)$, where $A := Q'_1\Sigma^{-1}_{-1,-1}U_{-1,-1}\Sigma^{-1}_{-1,-1}Q_1$ and $IG(\cdot,\cdot)$ denotes the Inverse Gamma $(\cdot,\cdot)$ distribution. Using this result and permutations, the partition can be done for the $p$ random variables in every step. Hence, this is not a standard Gibbs sampler because partitions change in every step; however, convergence can be established using results from Asmussen and Glynn (2011).

*The Khare-Rajaratnam family of flexible priors for decomposable graphs*

When $G$ is decomposable and its vertices are ordered according to a perfect elimination scheme (Khare and Rajaratnam, 2012), there exists a wider family of more flexible priors developed by Khare and Rajaratnam (2011). The parameter $\delta$ of the $IGW(U, \delta)$ family is common for all $D_{ii}$; however, for decomposable graphs a more flexible prior with pdf of the form $\bar{\pi}_{U,\delta}(\Sigma) \propto \exp\left(-\frac{1}{2}tr(\Sigma^{-1}U) - \sum_{i=1}^{p}\frac{\delta_i}{2}\log D_{ii}\right), \Sigma \in \mathbb{P}_G, U \in \mathbb{P}^+, \boldsymbol{\delta} = (\delta_1, \delta_2, \ldots, \delta_P)$ can be used. In this prior density, every $D_{ii}$ has its own shape parameter $\delta_i$. The price paid for this extra flexibility is that the graph $G$ has to be decomposable. When considering the modified Cholesky decomposition of the covariance matrix, the density in terms of $L$ and $D$ is:

$$\bar{\pi}_{U,\delta}(L, D) = \exp\left(-\frac{1}{2}tr((L')^{-1}D^{-1}L^{-1}U) - \sum_{i=1}^{p}\frac{\delta_i - 2n_i}{2}\log D_{ii}\right), L \in \mathcal{L}_G, D \in \mathcal{D}$$

Sufficient conditions for this to be a proper density are: $U \in \mathbb{P}^+$, $\delta_i > 2n_i + v_i + 2$ (Khare and Rajaratnam, 2011). This prior is conjugate because the posterior density, given by:

$$\pi(L, D|Y) \propto \exp\left(-\frac{1}{2}tr((L')^{-1}D^{-1}L^{-1}(U + NS)) - \sum_{i=1}^{p}\frac{N + \delta_i - 2n_i}{2}\log D_{ii}\right), L \in \mathcal{L}_G, D \in \mathcal{D},$$

is a $\bar{\pi}_{\tilde{U},\tilde{\delta}}(L, D)$ density, where $\tilde{U}$ is as defined before and $\tilde{\boldsymbol{\delta}}_{p \times 1} = \{\tilde{\delta}_i\} = N + \delta_i - 2n_i$. Hereinafter, this family of priors will be denoted as GWKR$(\boldsymbol{\delta}, U)$.

If in addition to be decomposable the graph is also homogeneous, direct sampling from the posterior can be performed (this case is discussed later), otherwise MCMC methods are used to draw samples from the posterior. Details of a block Gibbs sampler and the proof of its convergence can be found in Khare and Rajaratnam (2011). The full conditional distributions used in Khare and Rajaratnam's Gibbs sampler (Khare and Rajaratnam, 2011) are the following. Let $G = (V, E)$ be a decomposable graph with its vertices ordered according to a perfect elimination scheme, let $LDL'$ be the modified Cholesky decomposition of the covariance matrix $\Sigma$ and let $L^G_{\cdot v} = (L_{uv})_{u > v, (u,v) \in E}, v = 1, 2, \ldots, p - 1$. Then: $L^G_{\cdot v}|L \setminus L^G_{\cdot v}, D, Y \sim N(\mu^{v,G}, M^{v,G}) \forall v = 1, 2, \ldots, p - 1$, where

$$\mu^{v,G}_u = \mu^v_u + \sum_{u' > v:(u',v) \in E} \sum_{\substack{w > v:(w,v) \notin E \\ \text{or } w < v: (L^{-1})_{vw} = 0}} M^{v,G}_{uu'}\left(L^{-1}\tilde{U}(L')^{-1}\right)_{vv}\left((LDL')^{-1}\right)_{u'w}\mu^v_w \forall u > v, (u, v) \in E$$

$\mu^v_u = \frac{(L^{-1}\tilde{U})_{vu}}{(L^{-1}\tilde{U}(L')^{-1})_{vv}}$ $\forall u$ such that $(L^{-1})_{vu} = 0$, $((M^{v,G})^{-1})_{uu'} = \left(L^{-1}\tilde{U}(L')^{-1}\right)_{vv}\left((LDL')^{-1}\right)_{vv}$ $\forall u, u' > v, (u, v), (u', v) \in E$ and $D_{ii}|L, Y \sim IG\left(\tilde{\delta}_i/2, \left(L^{-1}\tilde{U}(L')^{-1}\right)_{ii}/2\right), i = 1, 2, \ldots, p$. In the

definition of $\mu^{v,G}$, notation $w: (L^{-1})_{vw} = 0$ refers to functional zeros, that is, $(L^{-1})_{vw}$ is zero as a function of the entries of $L$.

*Covariance graph models for homogeneous graphs*

For covariance graph models, certain properties of the graph $G = (V, E)$ have appealing mathematical consequences on the estimation problem. Covariance graph models take advantage of the fact that homogeneous graphs admit a Hasse ordering of their nodes (see Appendix A). The importance of having a graph with this sort of ordering is summarized in the following theorem (Khare and Rajaratnam, 2011): Let $G = (V, E)$, be a homogeneous graph with a Hasse ordering of its nodes. Then, $\Sigma = LDL' \in \mathbb{P}_G \Leftrightarrow L \in \mathcal{L}_G \Leftrightarrow L^{-1} \in \mathcal{L}_G$, that is, matrices $L$ and $L^{-1}$ preserve the pattern of zeros in $\Sigma$. When $G$ is homogeneous, direct samples from the posterior can be obtained by reparameterization in terms of $T = L^{-1}$. Let $\boldsymbol{x}_i := \{T_{ij}\}_{j<i,(i,j)\in E}$, then, it follows that the random vectors $\boldsymbol{x}_1, \boldsymbol{x}_2, \ldots, \boldsymbol{x}_{m-1}$ are mutually independent and distributed as follows $\boldsymbol{x}_i | D \sim MVN((U^{<i})^{-1}\boldsymbol{U}_{.i}^{\leq}, D_{ii}(U^{<i})^{-1})$. In addition, $D_{11}, D_{22}, \ldots, D_{pp}$ are also mutually independent with the following distribution $D_{ii} \sim IG\left(\frac{\delta - 2n_i - v_i}{2} - 1, \frac{U_{ii} - (\boldsymbol{U}_{.i}^{\leq})'(U^{<i})^{-1}\boldsymbol{U}_{.i}^{\leq}}{2}\right)$, where $v_i = |\{j: j < i, (i,j) \in E\}|$ and $U^{<i}, \boldsymbol{U}_{.i}^{\leq}$ and $U_{ii}$ form to the following submatrix of matrix $U$, $\begin{pmatrix} U^{<i} & \boldsymbol{U}_{.i}^{\leq} \\ (\boldsymbol{U}_{.i}^{\leq})' & U_{ii} \end{pmatrix}$, and are defined as follows: $U^{<i} = (U_{jk})_{j,k<i,(i,j),(i,k)\in E}$, and $\boldsymbol{U}_{.i}^{\leq} = (U_{ji})_{j<i,(i,j)\in E}$. From these conditional and marginal distributions, direct sampling can be performed.

**Adapting GCovGM to genome-wide prediction**

The linear regression model considered here is the following:
$$\boldsymbol{y} = W\boldsymbol{g} + \boldsymbol{e} \qquad (1)$$
where $\boldsymbol{y} \in \mathbb{R}^n$ is an observable random vector containing response variables (e.g., corrected phenotypes or deregressed BV), $\boldsymbol{g} \in \mathbb{R}^m$ is an unknown vector of marker allele substitution effects, $\boldsymbol{e} \in \mathbb{R}^n$ is a vector of residuals, $W_{n \times m}$ is a matrix whose entries correspond to one to one mappings from the set of genotypes to a subset of the integers for every individual at every locus $W = \{w_{ij}\} = \begin{cases} 1, if\ genotype = BB \\ 0, if\ genotype = BA \\ -1, if\ genotype = AA \end{cases}$, where $w_{ij}$ is map corresponding to the genotype of the $i^{th}$ individual for the $j^{th}$ marker. The distributional assumptions are: $\boldsymbol{g}|\Sigma \sim MVN(0, \Sigma)$ and $\boldsymbol{e}|\sigma^2 \sim MVN(0, \sigma^2 I)$ which implies $\boldsymbol{y}|\boldsymbol{g}, W, \sigma^2 \sim MVN(W\boldsymbol{g}, \sigma^2 I)$. In standard GCovGM, it is assumed that $N > 1$ independent and identically distributed multivariate Gaussian random variables are observed. On the other hand, in genome-wide prediction, most of the times phenotypic data corresponds to a single $n$-dimensional vector and the target is estimating $\Sigma$ instead of $Var[\boldsymbol{y}] := V = W\Sigma W' + \sigma^2 I$, that is, the target is estimating the covariance matrix of an unobservable $p$-dimensional random variable using a single $n$-dimensional vector containing phenotypes and the genomic information contained in $W$. Zhang et al. (2013) proposed methods to estimate covariance matrices corresponding to the sum of a low rank symmetric matrix and a sparse matrix, but these methods require a sample size larger than one and do not estimate $\Sigma$ directly. These are the difficulties encountered when considering the theory of GCovGM as a mean to model correlated marker effects in genome-wide prediction.

**A hierarchical Bayes formulation**

The flexibility of Hierarchical Bayesian modeling permits to cope with the problem of adapting GCovGM to genome wide prediction; it provides a simple and principled solution. The residual variance is given the following conjugate prior: $\sigma^2 \sim IG\left(\frac{a}{2}, \frac{b}{2}\right)$. Regarding the covariance matrix of marker effects, under the conventional GCovGM, Khare and Rajaratnam (2011) provided recursive equations to find the posterior mean in closed form for homogeneous graphs. However, as explained in the previous section, the target here is estimating the covariance matrix of an unobservable random vector; therefore, even for this sort of graphs sampling from the joint posterior distribution is required. To this end, the following simple but useful property permits the use of a Gibbs sampler. Notice that under model 1 it follows that $\pi(\Sigma|Else) = \pi(\Sigma|\boldsymbol{g})$. This property, and the conjugacy of the priors considered here (IGW and GWKR), imply that the full conditional of $\Sigma$ pertains to the same family of the prior. Therefore, because it is possible to obtain samples from these families and all other full conditionals are standard distributions, a Gibbs sampler can be implemented (Robert and Casella, 2010). Under the model termed Bayes GCov: $\Sigma|G \sim IGW(\delta, U)$ which can be used for general graphs. Then, the joint posterior has the following form:

$$\pi(\boldsymbol{g}, \sigma^2, \Sigma|\boldsymbol{y}, G) \propto (\sigma^2)^{-\frac{n}{2}} \exp\left(\frac{-1}{2\sigma^2}(\boldsymbol{y} - W\boldsymbol{g})'(\boldsymbol{y} - W\boldsymbol{g})\right) |\Sigma|^{-1/2} \exp\left(\frac{-1}{2}\boldsymbol{g}'\Sigma^{-1}\boldsymbol{g}\right)$$

$$\times \exp\left(-\frac{tr(\Sigma^{-1}U)}{2} - \frac{\delta}{2} \log|\Sigma|\right) (\sigma^2)^{-\left(\frac{b}{2}+1\right)} \exp\left(\frac{-a}{2\sigma^2}\right)$$

and $\Sigma|Else \sim IGW(\delta^*, U^*), U^* \coloneqq U + \boldsymbol{g}\boldsymbol{g}', \delta^* \coloneqq \delta + 1$. If $G$ is decomposable and the conditional prior for $\Sigma$ is a GWKR($\boldsymbol{\delta}, U$) distribution, then this variation of the model is referred to as Bayes GCov-KR. In this case, the full conditional distribution of $\Sigma$ is GWKR($\boldsymbol{\delta}^*, U^*$), where $\boldsymbol{\delta}^*_{m \times 1} \coloneqq \{\delta_i^*\} = 1 + \delta_i - 2n_i$. Finally, under the conditional prior $IGW(\delta, U)$, if the graph is homogeneous, the model is denoted as Bayes Cov-H just to stress the fact that this is a homogenous GCovGM and therefore, it is reparameterized in terms of the modified Cholesky decomposition of $\Sigma$. In this case, the Gibbs sampler is more efficient due to the fact that direct samples from the full conditional distribution of $\Sigma$ can be drawn.

**Some criteria to define $G$**

One of the first steps to carry out analyses with our models is defining the graph $G$, that is, defining the marginal covariance structure of marker allele substitution effects. To this end, some approaches based on genetic criteria are presented in this section. The first one is based on the idea of spatial correlation (Gianola et al. 2003, Yang and Tempelman 2012). Using a physical or linkage map, a window is defined based on a given map distance, or a given number of markers and it is slid across each chromosome. The order of markers is dictated by the physical or linkage map. This strategy induces a differentially banded or a banded covariance matrix. A second approach is based on the use of biological information. Using tools such as gene annotation information, markers can be clustered based on their function using approximations like the one presented in Peñagaricano et al. (2015). This will create groups or sets of loci and there are two options: permit correlations among effects of markers in different blocks or not. Finally, linkage disequilibrium between loci can be used. In this case, one of the metrics used to assess LD is chosen and those pairs of loci having a metric greater than a predefined threshold will be neighbors in $G$.

**Data analyses**

One of the main issues related to GCovGM is their computational burden. Consequently, to ensure computational tractability, two small datasets were simulated in order to implement the proposed models. A single genome formed by 5 chromosomes of 10 cM length each, with 1605 biallleic markers and 1000 biallelic QTL was simulated. This genome was created via a forward-in-time approach using software QMSim (Sargolzaei and Schenkel, 2013). Phenotypic records were created as the sum of the breeding value and an error term. For dataset 1, QTL effects were drawn from independent zero mean Gaussian distributions and were scaled such that the additive genetic variance was equal to 50. On the other hand, for dataset 2, QTL allele substitution effects were simultaneously drawn from a multivariate Gaussian distribution with null vector mean and a banded covariance matrix with bandwidth of size 10. These effects were then scaled in order to have an additive variance of 50. Residuals were drawn from independent Gaussian distributions with null mean and variance equal to 50, consequently, heritability was 0.5. Only markers with a minor allele frequency larger than 0.08 were considered in the analyses and ten replicates of each dataset were simulated. The graph $G$ was based on windows defined by a fixed number of marker loci (6), which induces a decomposable-non-homogeneous graph; therefore, models Bayes GCov-KR and Bayes GCov were fitted. Bayes A, a Bayesian model assuming uncorrelated effects, which is frequently used in genome-wide prediction, was also fitted. Training sets were formed by individuals from generations zero and one, and validation sets were comprised of individuals from generation 2.

Models were also fit to a real dataset. This dataset corresponded to a subset of the multibreed Angus-Brahman herd described in Elzo et al. (2012) and Elzo et al. (2013), containing 102 animals genotyped with the Illumina GoldenGate Bovine3K BeadChip (Illumina, Inc., 2011). Three traits were analyzed: daily feed intake (DFI, $\hat{h}^2 =0.31$), ultrasonic measure of percent of intra-muscular fat (UPIMF, $\hat{h}^2 = 0.53$) and body weight taken the same day than UPIMF (UW, $\hat{h}^2 = 0.54$) corrected for contemporary group (year-pen), age of dam, age of individual, sex, Brahman fraction and heterozygosity of the individual. The training set was defined as the oldest 72 individuals and the validation set as the remaining ones. The graph $G$ was defined using windows of 5 adjacent markers and only within chromosome correlations were allowed.

Predictive performance was assessed using the following criteria. Pearson correlation of phenotypes and predicted breeding values in the validation set (predictive ability) for real and simulated data, and the Pearson correlation between true and predicted breeding values (accuracy) in training and validation sets for simulated data. In each analysis, 15000 MCMC samples (first 5000 were considered burn in) were obtained using the Gibbs samplers described before. Analyses were performed using in-house R scripts (R Core Team, 2015).

## Results

Tables 1 and 2 present the summary of the performance of the models fitted to simulated and real datasets respectively.

**Table 1** Average (over 10 replicates) predictive abilities (APA), accuracies in training (AAT) and validation (AAV) sets for simulated datasets 1 and 2 (standard deviations in brackets).

| Model | Dataset 1 | | | Dataset 2 | | |
|---|---|---|---|---|---|---|
| | APA | AAT | AAV | APA | AAT | AAV |
| Bayes GCov | 0.432 | 0.739 | 0.573 | 0.432 | 0.716 | 0.557 |
| | (0.075) | (0.056) | (0.078) | (0.128) | (0.038) | (0.099) |
| Bayes GCov-KR | 0.441 | 0.740 | 0.573 | 0.444 | 0.743 | 0.566 |
| | (0.071) | (0.058) | (0.075) | (0.094) | (0.056) | (0.089) |
| Bayes A | 0.352 | 0.684 | 0.417 | 0.404 | 0.711 | 0.526 |
| | (0.161) | (0.064) | (0.081) | (0.048) | (0.051) | (0.123) |

**Table 2** Predictive abilities for real data analyses

| Model | DFI | UPIMF | UW |
|---|---|---|---|
| Bayes GCov | 0.019 | 0.036 | 0.078 |
| Bayes GCov-KR | 0.027 | 0.032 | 0.068 |
| Bayes A | 0.081 | -0.006 | -0.007 |

According to all criteria (APA, AAT, AAV), our models clearly outperformed Bayes A in the two simulated datasets, the differences being more marked in the case of independent QTL effects (dataset 1). In these datasets, the flexibility of the GWKR priors yielded a better predictive performance. Also, the performance of our methods tended to be less variable; Bayes A showed a smaller variation only for APA in dataset 2. In the real data set, all predictive abilities were low. For DFI, Bayes A had the highest predictive ability, while for UPIMF and UW it had the lowest. For these two traits, Bayes GCov had the best predictive performance.

## Discussion

**General comments about the models**

In this study, the theory of GCovGM was adapted to genome-wide prediction through hierarchical Bayesian modeling. This development permits to account for correlated marker allele substitution effects in a flexible way. This flexibility is due to the ability of our models to accommodate covariance structures arising from biological considerations such as information from metabolic pathways and not only from the assumption of spatial correlation as has been done in previous studies (Gianola et al., 2003; Yang and Tempelman, 2012). Thus, covariances between effects of markers which are not physically linked are permitted. Furthermore, the possibility of defining the graph $G$ using tools such as gene annotation provides a way to incorporate biological information in the prediction process.

Several approaches to define the graph based on biological principles were presented. These approaches involve the assumption of spatial correlation and the aforementioned use of existing bioinformatics tools to create "functional" sets of SNP whose effects are correlated. In general, the second strategy would induce graphs with no special properties. However, due to the theoretical and numerical advantages of decomposable graphs discussed previously, it is convenient to work with this sort of graphs whenever possible. To this end, in a submitted paper, we have proven two propositions and a corollary that provide conditions on the edges set and the ordering of markers, such that the induced graph is decomposable. For the sake of completeness, these propositions and the corollary are presented in Appendix B. Proposition 1 in Appendix B is the most general, but when $G$ is defined using biological information and subsets of different "functional" SNP sets are allowed to be correlated, its conditions are more difficult to satisfy. On the other hand, proposition 2 in Appendix B and its corollary are more restrictive in terms of the covariance structure, but they provide easier ways to order markers and define the edge set, that guarantee decomposability. Once the "functional" sets have been defined, if these conditions are not satisfied, these theoretical results provide a basis to find a decomposable super-graph containing the original graph $G$. Such a super-graph is known as the cover of $G$ (Khare and Rajaratnam, 2012).

In GCovGM, the family of homogeneous graphs is the one with more attractive properties. This is why the implementation of Bayes GCov-H is easier and faster because direct sampling of $\Sigma$ is feasible. However, finding this kind of graphs is, in general, not an easy problem. An example of a homogeneous graph is a rooted tree where all nodes are children of a single parent (the root). Thus, under the approach of using biological information to define the graph $G$, a homogeneous graph can easily be found as follows: The tree structure mentioned above is imposed to each "functional" set and no correlations between effects of markers in different sets are allowed. It also holds when each "functional" set is assumed to be a complete. All the strategies mentioned before might appear restrictive, but notice that assuming independent marker effects amounts to imposing a covariance structure as well. In fact it is a special case of our approach when the edge set is the empty set.

Here, the focus was on Bayesian models because under the GCovGM framework, they can deal with the "big p small n" setting. However, in Appendix C, a frequentist approach to find the empirical BLUP of $\boldsymbol{g}$ is presented. This formulation is based on the EM algorithm (Dempster et al., 1977) combined with GCovGM theory which permits obtaining estimators of dispersion parameters $\Sigma$ and $\sigma^2$ which are used to build the mixed model equations corresponding to model 1 whose solution yields the empirical BLUP of $\boldsymbol{g}$ (Henderson,1963). This formulation involves a partition of data induced by the assumption that different groups (e.g., half-sib families) have different sets of marker effects. Such an assumption has also been considered by other authors like Gianola et al. (2003).

Even with the aid of bioinformatics, biochemistry and physiology to construct the graph $G$, it may not reflect the actual underlying covariance structure, but important correlations might be captured resulting in an improvement of the accuracy of genome-wide prediction. Covariance model selection involves finding the pattern of zeros and estimating the non-zero elements of either the precision or the covariance matrix (Bickel and Levina 2008; Khare et al., 2013). Model selection in GCovGM has not been as well studied as its counterpart in Gaussian concentration graph models. There exist some frequentist methods that induce sparsity based on penalized likelihood approaches (Bien and Tibshirani, 2011) and others based on the idea of inducing sparsity in the parameter $L$ of the modified Cholesky decomposition of $\Sigma$ (Rothman et al., 2010). From the Bayesian perspective, some methods based on the Bayesian lasso have been proposed, e.g., Wang (2012), but their main limitation is the computational burden. Some of these models could be implemented in genome-wide prediction following approaches similar to those presented in this study.

**Extension to multiallelic loci**

Here, biallelic loci were considered, but in some cases multiallelic loci have to be dealt with. In the future, models could be fit using genotypes for actual genes instead of molecular markers. In such a case, there could be more than two alleles per locus. A similar situation occurs when fitting effects of haplotypes built from two or more consecutive markers (Meuwiseen et al., 2001; Calus et al., 2008). The methods developed here can be easily extended to the multiallelic case. If there are $a_k$ alleles at locus $k$, then the corresponding columns of the design matrix are formed by defining $a_k - 1$ variables as follows: $W^k = \{w_{ij}^k\} = \begin{cases} 1, if\ genotype = A_j A_j \\ 0, if\ genotype = A_j -\ , j = 1,2,\dots, a_k - 1 \\ -1, if\ genotype = -- \end{cases}$, where $w_{ij}^k$ is the genotype of the $i^{th}$ individual for the $j^{th}$ allele at locus $k$ and "$-$" represents an allele different from $A_j$. The graph $G$ can be built based on the ideas discussed before, with extra considerations at the intra-locus level. For example, it could be assumed that effects of alleles of the same locus are all correlated.

**Data analyses**

In general, Bayes GCov and Bayes GCov-KR outperformed Bayes A except for DFI. Differences between our models and Bayes A were more marked when QTL effects were independent and models considered SNP effects (dataset 1). In the ideal scenario where the model considers the causal variants (the QTL) instead of markers, the benefit of accounting for marginal correlation is smaller as suggested by a smaller difference in the three criteria used to assess predictive performance. This behavior may suggest that when considering the causal variants instead of proxies as the SNP, models assuming independent effects yield an acceptable predictive performance even when the true covariance matrix is non-diagonal. Hopefully, this ideal scenario where the causal variants determining a phenotype, or at least most of them, are known will be reached in the near future. The largest difference between the method Bayes A and our methods (15.7%) was observed for AAV in dataset 1, while the smallest one (0.5%) was observed for AAT in dataset 2, in both cases, when comparing it with Bayes GCov-KR. Although Bayes GCov-KR had higher APA, AVA and ATA values in the simulated datasets, notice that the differences compared to Bayes G-Cov were small, being slightly larger in dataset 2; therefore, in these simulations the gain in fitting a more complex model which considers as many shape parameters as markers did not yield a notorious gain in accuracy or predictive ability. The gains in accuracy in the validation set observed in dataset 1 are larger than those found by Yang and Tempelman (2012) when comparing their antedependence models with their independent marker effects counterparts Bayes A and Bayes B, whereas gains in accuracy observed in dataset 2 were comparable (they found increments in accuracy of breeding values in the testing population up to 3%). The simulated data in Yang and Tempelman (2012) were similar to dataset 1, where it is expected that correlation among SNP arises from physical proximity to the same causal variants. They also considered a heritability value of 0.5. As to the real data analyses, first of all it has to be considered that these analyses are just a proof of concept and analyses involving larger populations with denser marker panels are needed. We did not consider a larger population and a denser SNP panel due to limited computational resources. For DFI, Bayes A had a higher PA whereas for UPIMF and UW Bayes GCov had the best PA followed by Bayes GCov-KR. For a mice population, Yang and Tempelman (2012) also found that a model assuming independence (Bayes B) outperformed one of their models accounting for correlation (ante-Bayes A) in terms of predictive ability. Although these traits have medium to high estimated heritabilities (Elzo et al., 2012; Elzo et al., 2013) PA were very low. Perhaps this is due to the low number of SNP considered here (2407). The study of Gianola et al. (2003) did not consider data analysis.

## Final remarks

This paper introduces the theory of GCovGM in the context of genome-wide prediction which permits to account for correlated marker effects in a very flexible way in terms of the marginal covariance structure. Models developed here also allow incorporating biological information in the prediction process through its use when building graph $G$.

## Acknowledgements

C.A. Martínez thanks Fulbright Colombia and "Departamento Adiministrativo de Ciencia, Tecnología e Innovación" COLCIENCIAS for supporting his PhD and Master programs at the University of Florida through a scholarship, and Bibiana Coy for her support and constant encouragement.

## References

Asmussen, S., & Glynn, P.W. (2011). A new proof of convergence of MCMC via the ergodic theorem. *Statistics and Probability Letters, 81*, 1482-1485.

Bickel, P.J., & Levina, E. (2008). Covariance regularization by thresholding. *The Annals of Statistics, 36(6)*, 2577-2604.

Bien, J. & Tibshirani, R.J.. (2011). Sparse estimation of a covariance matrix. *Biometrika, 98(4),* 807-820.

Calus M.P.L., Meuwissen, T.H.E., de Roos, A.P.W., & Veerkamp, R.F. (2008). Accuracy of genomic selection using different methods to define haplotypes. *Genetics, 178,* 553-561.

Carvalho, C.M., Massam, H., & West, M. (2007). Simulation of hyper-inverse- Wishart distributions in graphical models. *Biometrika, 94*, 647-659.

Dempster, A.P., Laird, N.M., & Rubin, D.B. (1977). Maximum likelihood from incomplete data via the EM algorithm. *Journal of the Royal Statistical Society Series B, 39(1)*, 1-38.

Diaconis, P., & Ylvisaker, D. (1979). Conjugate priors for exponential families. *The Annals of Statistics, 7(2)*, 269-281.

Elzo, M. A., Lamb, G.C., Johnson, D.D., Thomas, M.G., Misztal, I., Rae, D.O., Martinez, C.A., Wasdin, J.G., & Driver, J.D. (2012). Genomic-polygenic evaluation of Angus-Brahman multibreed cattle for feed efficiency and postweaning growth using the Illumina3k chip. *Journal of Animal Science, 90*, 2488-2497.

Elzo, M. A., Martinez, C.A., Lamb, G.C., Johnson, D.D., Thomas, M.G., Misztal, I., Rae,D.O., Wasdin, J.G., & Driver, J.D.. (2013). Genomic-polygenic evaluation for ultrasound and weight traits in Angus-Brahman multibreed cattle with the Illumina3k chip. *Livestock Science, 153*, 39-49.

Gianola, D., Perez-Enciso, M., & Toro, M.A. (2003). On marker-assisted prediction of genetic value: Beyond the Ridge. *Genetics, 163*, 347-365.

Henderson, C.R. (1963). Selection index and expected genetic advance. In W.D. Hanson & H.F. Robinson (Eds.), *National Academy of Sciences-National Research Council,* publication 982.

Illumina, Inc., 2011. GoldenGate Bovine3K Genotyping BeadChip. Illumina Data Sheet. San Diego, CA. www.illumina.com/Documents//products/datasheets/datasheet_bovine3k.pdf.

Khare, K., & Rajaratnam, B. (2011). Wishart distributions for decomposable covariance graph models. *The Annals of Statistics, 39(1)*, 514-555.

Khare, K., & Rajaratnam, B. (2012). Sparse matrix decompositions and graph characterizations. *Linear Algebra and its Applications, 437*, 932-947.

Khare, K., Oh, S., & Rajaratnam, B. (2015). A convex pseudo-likelihood framework for high dimensional partial correlation estimation with convergence guarantees. *Journal of the Royal Statistical Society Series B, 77(4)*, 803-825.

Letac, G., & Massan, H. (2007). Wishart distributions for decomposable graphs. *The Annals of Statistics, 35(3),* 1278-1323.

Meuwissen, T.H.E., Hayes B.J., & Goddard, M.E. (2001). Prediction of total genetic value using genome-wide dense marker maps. *Genetics, 157*,1819-1829.

Peñagaricano, F., Weigel, K.A., Rosa, G.J.M., & Kathib, H. (2013). Inferring quantitative trait pathways associated with bull fertility from a genome-wide association study. *Frontiers in Genetics, 3*, 307.

R Core Team (2015). R: A language and environment for statistical computing. R foundation for statistical computing, Vienna, Austria. *URL https://www.R-project.org/.*

Rajaratnam, B., Massam, H., & Carvalho, C. (2008). Flexible covariance estimation in graphical Gaussian models. *The Annals of Statistics, 36(6)*, 2818-2849.

Rothman, A., Levina, E., & Zhu, J. (2010). A new approach to Cholesky-based covariance regularization in high dimensions. *Biometrika, 97(3),* 539-550.

Sargolzaei, M., & Schenkel, F.S. (2013). *QMSim User's Guide Version 1.10*. Centre for Genetic Improvement of Livestock, Department of Animal and Poultry Science, University of Guelph, Guelph, Canada.

Silva, R., & Ghahramani, Z. (2009). The Hidden Life of Latent Variables: Bayesian Learning with Mixed Graph Models. *Journal of Machine Learning Research, 10*, 1187-1238.

Stein, C. (1975). Estimation of a covariance matrix. In *Reitz lecture*. 39[th] annual meeting. IMS Atlanta, GA

Wang, H. (2012). Bayesian Graphical Lasso models and efficient posterior computation. *Bayesian Analysis, 7(4)*, 867-866.

Yang, W., & Tempelman, R.J. (2012). A Bayesian Antedependence Model for Whole Genome Prediction. *Genetics*, *190*, 1491-1501.

Zhang, L., Sarkar, A., & Mallick, B.K. (2013). Bayesian low rank and sparse covariance matrix decomposition. *arXiv preprint,* arXiv:1310.4195.

## Appendix A: Basic Concepts in Graph Theory

**Undirected graph.** An undirected graph $G$ is defined as a collection of two objects $G = (V, E)$ where $V$ is the set of vertices (finite) and $E \subseteq V \times V$ is the set of edges satisfying:
$$(u, v) \in E \Leftrightarrow (v, u) \in E.$$
**Neighbor vertices**. Let $G = (V, E)$ be an undirected graph. The vertices $u, v \in V$ are said to be neighbors if $(u, v) \in E$.

**P-path**. A p-path is a collection of p distinct vertices $u_1, u_2, \ldots, u_p$ such that $(u_i, u_{i+1}) \in E, i = 1, 2, \ldots, p-1$, that is, $(u_i, u_{i+1})$ are neighbors for $i = 1, 2, \ldots, p-1$.

**P-cycle**. A p-cycle is a collection of p distinct vertices $u_1, u_2, \ldots, u_p$ such that $(u_i, u_{i+1}) \in E, i = 1, 2, \ldots, p-1$ and $(u_p, u_1) \in E$

**Clique**. A subset $V_0 \subset V$ is a clique if $(u, v) \in E \ \forall \ u, v \in V_0$.

**Maximal clique**. A subset $V_0 \subset V$ is defined to be a maximal clique if $V_0$ is a clique and there does not exist a clique $\bar{V}$ such that $V_0 \subset \bar{V} \subseteq V$.

**Ordered graphs**. Let $G = (V, E)$ and let $\sigma$ be an ordering of $V$, that is, a bijection from $V$ to $\{1, 2, \ldots, |V|\}$. Then, the ordered graph $G_\sigma = (\{1, 2, \ldots, |V|\}, E_\sigma)$ is defined as follows: $(i, j) \in E_\sigma$ iff $\left(\sigma^{-1}(i), \sigma^{-1}, (j)\right) \in E$.

**Perfect elimination ordering**. An ordering $\sigma$ of a graph $G = (V, E)$ is defined to be a perfect elimination ordering if a triplet $\{i, j, k\}$ with $i > j > k$ such that $(i, j) \notin E_\sigma$ and $(i, k), (j, k) \in E_\sigma$ does not exist.

**Subgraph.** The graph $G' = (V', E')$ is said to be a subgraph of graph $G = (V, E)$ if $V' \subseteq V$ and $E' \subseteq E$.

**Induced subgraph.** Consider the graph $G = (V, E)$ and a subset $A \subseteq V$. Define $E_A = (A \times A) \cap E$. The subgraph $G_A = (A, E_A)$ is defined to be a subgraph of $G$ induced by $A$. **Decomposable graph**. An undirected graph $G = (V, E)$ is a decomposable graph if it does not contain a cycle of length greater than or equal to four as an induced subgraph. It turns out that decomposable graphs are characterized by the existence of a perfect elimination ordering of their vertices; therefore, a graph $G = (V, E)$ is decomposable iff its vertices admit a perfect elimination ordering.

**Connected graph**. A graph $G$ is said to be connected if any pair of distinct vertices in $G$ are connected, that is, there exists a path between them.

**Directed edges**. An edge is said to be directed if $(u, v) \notin E$ whenever $(v, u) \in E$. If $(v, u)$ is a directed edge then $v$ is said to be a *parent* of $u$ and $u$ is said to be a *child* of $v$.

**Directed graph**. A graph $\mathcal{D} = (V, E)$ such that its edges are directed is defined as a directed graph.

**Directed acyclic graph**. A directed acyclic graph (DAG) is a directed graph with no cycles.

**Tree**. A tree is a connected graph with no cycle of length greater or equal than 3.

**Rooted tree**. A rooted tree is a tree in which a particular node is distinguished from the others and designated the root of the tree. This node is the ancestor of all other nodes in the tree. An ancestor of a node $u$ in a rooted tree with root node $r$ is any node in the path from $r$ to $u$.

**Homogeneous graph.** An undirected graph $G = (V, E)$ is defined to be homogeneous if for all $(u, v) \in E$, either:
$$\{i: i = u \text{ or } (i, u) \in E\} \subseteq \{i: i = v \text{ or } (i, v) \in E\}$$
or
$$\{i: i = v \text{ or } (i, v) \in E\} \subseteq \{i: i = u \text{ or } (i, u) \in E\}.$$
An equivalent definition is the following. A graph $G = (V, E)$ is said to be homogeneous if it is decomposable and it does not have a 4-path as an induced subgraph. Homogeneous graphs have an equivalent representation in terms of directed rooted trees called Hasse diagrams.

**Hasse diagram**. A Hasse diagram is built as follows. For $i \in V$, let $\mathcal{N}(u) := \{i: i = u \text{ or } (i, u) \in E\}$. Whenever $\mathcal{N}(u) \subseteq \mathcal{N}(v)$ we write $v \to u$. If $u \to v$ and $v \to u$ it is said that there is a equivalence relation between $u$ and $v$. Using this relation, equivalence classes are created. For example, if $\mathcal{N}(u) = \mathcal{N}(v)$, then $u$ and $v$ are in the same equivalence class. The equivalence classes are the nodes of the Hasse diagram, formally, if $\bar{u}$ denotes the equivalence class containing node $u$, then the Hasse diagram of $G$ is a directed graph with node set $V_H := \{\bar{u}: u \in E\}$. The edge set $E_H$ is defined as follows. If $\bar{u} \neq \bar{v}$, $u \to v$, and $\nexists \, k$ such that $u \to k \to v$ then put a directed edge from $u$ to $v$.

**Hasse perfect vertex elimination scheme or Hasse ordering**. Once the Hasse diagram of $G$ has being built, the nodes of $G$ are ordered in the following way. The ordering is descending starting from the root of the tree; therefore, nodes pertaining to equivalence classes on the top of the Hasse diagram are assigned the largest levels. Within every equivalence class with more than one node, the ordering is arbitrary. Hence, the ordering is not unique. Any ordering that gives an ancestor a higher level than any of its descendants in the Hasse diagram of $G$ is defined to be a Hasse perfect vertex elimination scheme or simply a Hasse ordering of the nodes of $G$.

## Appendix B: Maximum likelihood estimation in covariance graph models

**Maximum likelihood estimation of Σ for general graphs, standard problem**

If the sample size $N$ is larger than $p$, then maximum likelihood estimation of $\Sigma$ is feasible. After removing constant terms from the negative log-likelihood the following is the objective function to be minimized: $l^*(\Sigma) = tr(\Sigma^{-1} S) + \log|\Sigma|, \Sigma \in \mathbb{P}_G$, where $S$ is the sample covariance matrix. Notice that the objective involves $\Sigma^{-1}$ instead of $\Sigma$. This objective function is not convex, which makes this minimization more difficult than the minimization problem for concentration graph models. One important feature of covariance graph models is that they correspond to curved exponential families instead of the well-studied exponential families as is the case of concentration graph models (Khare and Rajaratnam 2011), it poses a more challenging problem.

An iterative conditional fitting (ICF) algorithm to minimize $l^*(\Sigma)$ was developed by Chaudhuri et al. (2007); however, because we are dealing with a non-convex optimization problem, convergence to a global or even a local minimum is not guaranteed.

The algorithm is based on the following partition of $\Sigma$:

$$\Sigma = \begin{bmatrix} \Sigma_{11} & \Sigma'_{.1} \\ \Sigma_{.1} & \Sigma_{-1,-1} \end{bmatrix} \qquad (B.1)$$

where $\Sigma_{11}$ is the 1,1 entry of $\Sigma$, $\Sigma_{.1}$ is the first column of $\Sigma$ without the first entry and $\Sigma_{-1,-1}$ is the submatrix of $\Sigma$ resulting from deleting its first row and column. Using the standard rules for inversion by partitioning:

$$\Sigma^{-1} = \begin{bmatrix} \dfrac{1}{\gamma_1} & \dfrac{-\Sigma'_{.1}\Sigma^{-1}_{-1,-1}}{\gamma_1} \\ \dfrac{-\Sigma^{-1}_{-1,-1}\Sigma'_{.1}}{\gamma_1} & \Sigma^{-1}_{-1,-1} + \dfrac{\Sigma^{-1}_{-1,-1}\Sigma_{.1}\Sigma'_{.1}\Sigma^{-1}_{-1,-1}}{\gamma_1} \end{bmatrix}$$

where $\gamma_1 = \Sigma_{11} - \Sigma'_{.1}\Sigma^{-1}_{-1,-1}\Sigma_{.1}$. Notice that knowing $\Sigma$, we can get $(\Sigma_{.1}, \Sigma_{-1,-1}, \gamma_1)$ and vice versa; consequently, we have a one to one transformation. By using permutations, the same partition can be performed for every one of the $p$ random variables represented in graph $G$. The high level of the algorithm is the following:

1) Partition $\Sigma$ as $(\Sigma_{.1}, \Sigma_{-1,-1}, \gamma_1)$, 2) minimize $l^*(\Sigma)$ with respect to $\Sigma_{.1}$ treating as fixed the current values of $\Sigma_{-1,-1}$ and $\gamma_1$ and 3) minimize $l^*(\Sigma)$ with respect to $\gamma_1$ fixing the current values of $\Sigma_{.1}$ and $\Sigma_{-1,-1}$. The same is repeated for the $p$ variables and it corresponds to one iteration of the algorithm. The minimization problem is solved by minimizing the following quadratic form with respect to $\boldsymbol{\beta}_1$ (Chaudhuri et al.,2007):

$$\frac{-1}{\gamma_1}\left(2\boldsymbol{\beta}'_1 Q'_1 \Sigma^{-1}_{-1,-1} S_{.1} - \boldsymbol{\beta}'_1 Q'_1 \Sigma^{-1}_{-1,-1} S_{-1,-1} \Sigma^{-1}_{-1,-1} Q_1 \boldsymbol{\beta}_1 \right)$$

where $\boldsymbol{\beta}_1 := (\Sigma_{1j})_{(1,j)\in E}$, $S_{.1}$ and $S_{-1,-1}$ are elements obtained after partitioning $S$ as $\Sigma$ was partitioned in $(B.1)$ and $Q_1$ is a matrix of zeros such that: $\Sigma_{.1} = Q_1 \boldsymbol{\beta}_1$. This is a standard problem and its solution is $\widehat{\boldsymbol{\beta}}_1 = \left(Q'_1 \Sigma^{-1}_{-1,-1} S_{-1,-1} \Sigma^{-1}_{-1,-1} Q_1\right)^{-1} Q'_1 \Sigma^{-1}_{-1,-1} S_{.1}$. On the other hand, the solution to the second minimization is: $\hat{\gamma}_1 = S_{11} - 2\Sigma'_{.1}\Sigma^{-1}_{-1,-1}S_{.1} + \Sigma'_{.1}\Sigma^{-1}_{-1,-1}S_{-1,-1}\Sigma^{-1}_{-1,-1}\Sigma_{.1}$. Kauermann (1996) proposed to modify the objective function in order to make it a function of $\Sigma$, which makes the problem convex. The new objective function has the form: $\tilde{l}(\Sigma) = tr(\Sigma S^{-1}) + log|\Sigma|, \Sigma \in \mathbb{P}_G$. A trick based on using the maximal cliques of $G$ is applied to solve this problem and the solution is known as the Kauermann's dual estimator. Under certain conditions, convergence of the ICF algorithm at least to a local stationary point can be proved (Drton et al., 2006).

**Maximum likelihood estimation of $\Sigma$ for homogenous graphs, standard problem**

Recall that for homogeneous graphs, $\Sigma = LDL' \in \mathbb{P}_G \Leftrightarrow L \in \mathcal{L}_G \Leftrightarrow L^{-1} \in \mathcal{L}_G$; therefore, the objective function is written in terms of $(L, D)$ instead of $\Sigma$. Also, recall that after removing constant terms from the negative log-likelihood we get: $l^*(\Sigma) = tr(\Sigma^{-1}S) + log|\Sigma|, \Sigma \in \mathbb{P}_G$. The bijection from $\mathbb{P}_G$ to $\mathcal{L}_G \times \mathcal{D}$ induces:

$$l^*(L, D) = tr((L')^{-1} D^{-1} L^{-1} S) + log|D|, L \in \mathcal{L}_G, D \in \mathcal{D}$$

reparameterization in terms of $T = L^{-1}$ yields:

$$l^*(L, D) = tr(T' D^{-1} T S) + log|D|, T \in \mathcal{L}_G, D \in \mathcal{D}$$

$$= \sum_{i=1}^{p} \frac{1}{D_{ii}}(T_{i.} S T'_{i.}) + log\, D_{ii} \qquad (B.2)$$

where $T_{i.}$ Is the $i^{th}$ row of $T$.

To obtain the MLE of $\Sigma$, every summand in $(B.2)$ is minimized with respect to $D_{ii}$ and $T_{i.}$. Define $\boldsymbol{x}_i := \{T_{ij}\}_{j<i,(i,j)\in E}$; $N^<(i) := \{j: j < i, (i,j) \in E\}$ and construct the following matrix from the sample covariance matrix:

$$S_i = \begin{pmatrix} S^{<i} & S_{.i}^{<} \\ (S_{.i}^{<})' & S_{ii} \end{pmatrix} \quad (B.3)$$

where $\boldsymbol{S}_{.i}^{<} = (S_{ki})_{k<i,(i,k)\in E}$, $S^{<i} = (S_{kl})_{k,l\in N^<(i)}$. Then, the MLE are:

$$\hat{\boldsymbol{x}}_i = (S^{<i})^{-1} \boldsymbol{S}_{.i}^{<}$$
$$\hat{D}_{ii} = S_{ii} - (\boldsymbol{S}_{.i}^{<})'(S^{<i})^{-1} \boldsymbol{S}_{.i}^{<}$$

Combining all $\hat{D}_{ii}$ and $\hat{\boldsymbol{x}}_i$ we can build $\hat{D}$ and $\hat{T}$ and using them we have $\hat{\Sigma} = \hat{L}\hat{D}\hat{L}'$.

**Maximum likelihood estimation of $\Sigma$ in genome-wide prediction**

Unlike the Bayesian approach, envisaging a frequentist solution to the problem of adapting GCGM to genome-wide prediction under the model presented in the manuscript is not straightforward and we could not find a direct and principled method to cope with this problem. Therefore, some *ad hoc* extra assumptions were done in order to provide a frequentist formulation. The method proposed here involves two steps. The first one combines the EM algorithm (Dempster et al., 1977) with GCovGM to estimate covariance components. The second one involves plugging these estimates into mixed model equations corresponding to model 1 in order to obtain the empirical BLUP of $\boldsymbol{g}$ (Henderson, 1963).

According to the rationale of the EM-algorithm, we define $\boldsymbol{g}$ as the augmented or missing data, then we find the maximizers of the complete likelihood as if $\boldsymbol{g}$ were observable and finally we compute their expected values with respect to the distribution of the missing or augmented data given the observed data. As mentioned in the manuscript, maximum likelihood estimation of $\Sigma$ is only possible if $N > m$. In model 1, we have a single $n$-dimensional vector $\boldsymbol{y}$ and the target is to estimate the residual variance and the covariance matrix of the $m$-dimensional vector $\boldsymbol{g}$; therefore, in terms of the standard problem: $N=1$. Thus, an *ad hoc* solution is to assume that data can be split into $f > m$ groups such that each group has a different vector of marker effects, that is, $\boldsymbol{y}_i = W_i \boldsymbol{g}_i + \boldsymbol{e}_i, \forall i = 1,2,...,f$. Currently, as more and more animals are genotyped, for SNP panels of moderate density (e.g., 50K) the case $n > m$ can be found. For example, this is the case of the Holstein population in the US. However, this is not the most common situation and it is important to notice that it does not imply that $f > m$ which is the necessary condition to carry out maximum likelihood estimation of $\Sigma$. One of the simplest ways to split a population into $f$ groups is by considering families (e.g., half-sibs or full-sibs) as in Gianola et al. (2003). Currently, the requirement $f > m$ will be met by very few populations when considering a relatively small number of markers and this is the reason for not considering the frequentist approach in the manuscript. Notwithstanding, in this appendix we provide an approach to carry out maximum likelihood estimation of the dispersion parameter $\boldsymbol{\theta} := (\Sigma, \sigma^2)$ in a genome-wide prediction model based on multiple linear regression which later permits to obtain the empirical BLUP of $\boldsymbol{g}$.

It is also assumed that: $g_1, \ldots, g_f$ are iid $MVN(0, \Sigma)$, $e_1, \ldots, e_f$ are iid $MVN(0, \sigma^2 I_{n_i})$ and $Cov(g_i, e_{i\prime}) = 0, \forall\ 1 \leq i, i' \leq f$, where $n_i$ is the number of observations in group $i$; therefore, $\sum_{i=1}^{f} n_i = n$. Under these assumptions, the complete log-likelihood can be written as:

$$l(\sigma^2, \Sigma) = constants - \frac{n}{2}\log\sigma^2 + \frac{f}{2}\left(\log|\Sigma^{-1}| - tr(\Sigma^{-1}S_g)\right) - \frac{\|y - W^*g^*\|_2^2}{2\sigma^2} \quad (B.4)$$

$$S_g = \frac{1}{f}\sum_{i=1}^{f} g_i g_i', \quad g^* := (g_1' \cdots g_f')', \quad W^* = Block\ Diag.\{W_i\}_{i=1}^{n}.$$

The expected values of sufficient statistics for the covariance parameters taken with respect to the conditional distribution of the missing data given the observed data have to be found. The sufficient statistic for $\theta$ is $(S_g, e^{*\prime}e^*)$, $e^* = y - W^*g^*$. Also, given $y$, $g_1, \ldots, g_f$ are independent with the following distributions: $g_i|y_i \sim MVN\left(K_i^{-1}\frac{W_i'y_i}{\sigma^2}, K_i^{-1}\right)$, where $K_i := \frac{W_i'W_i}{\sigma^2} + \Sigma^{-1}$. Similarly, it follows that $e^*|y \sim MVN(\sigma^2 V^{-1}y, \sigma^2(I - \sigma^2 V^{-1}))$, where $V = W^{*\prime}I_f \otimes \Sigma W^* + R$. Hence,

$$E[S_g|y] = \frac{1}{f}\sum_{i=1}^{f} K_i^{-1}\left[I_m + \frac{1}{(\sigma^2)^2}W_i'y_i y_i'W_i K_i^{-1}\right] \quad (B.5)$$

$$E[e^{*\prime}e^*|y] = \sigma^2(n - \sigma^2 tr(V^{-1}) + \sigma^2 y'V^{-1}V^{-1}y) \quad (B.6)$$

Applying the Woodbury's identity, $E[S_g|y]$ can be alternatively expressed as:

$$E[S_g|y] = \frac{1}{f}\Sigma\left\{fI_m - \left[\sum_{i=1}^{f} W_i'V_i^{-1}(I_{n_i} - y_i y_i'V_i^{-1})W_i\right]\Sigma\right\} \quad (B.7)$$

where $V_i := W_i \Sigma W_i' + \sigma^2 I_{n_i}$. It does not require inversion of $\Sigma$, it requires inverting $f\ n_i \times n_i$ matrices. The expectation step of this EM algorithm consists of using either B.5 or B.7 to compute $E[S_g|y]$ and B.6 to compute $E[e^{*\prime}e^*|y]$, the maximization step is the one involving GCovGM. At iteration $t$, the maximization step involves the following computations:

$$(\hat{\sigma}^2)^{(t+1)} = \frac{\hat{q}^{(t)}}{n}, \hat{q}^{(t)} := E[e^{*\prime}e^*|y]\Big|_{\theta = \theta^{(t)}}$$

$$\hat{\Sigma}^{(t+1)} = h\left(\hat{S}_g^{(t)}\right), \hat{S}_g^{(t)} := E[S_g|y]\Big|_{\theta = \theta^{(t)}}$$

where $\hat{\Sigma}^{(t+1)}$ is computed using methods explained before. For homogeneous graphs, function $h(\cdot)$ has closed forms after reparameterizing the objective function in terms of $(T, D)$ as shown previously in this section. Once the algorithm converges and the maximum likelihood estimates of $\Sigma$ and $\sigma^2$ are obtained, these are plugged in the mixed model equations corresponding to model 1 to obtain the empirical BLUP of $g$ (Henderson, 1963):

$$\hat{g} = \left(W'W + \hat{\sigma}^2 \hat{\Sigma}^{-1}\right)^{-1} W'y.$$

**References (only those not included in the manuscript are presented here)**

Chaudhuri, S., Drton, M., & Richardson, T.M. (2007). Estimation of a Covariance Matrix with Zeros. *Biometrika, 94(1),* 199-216.

Drton, M., Eichler, M., & Richardson, T.S. (2006). Computing Maximum Likelihood Estimates in Recursive Linear Models with correlated Errors. *ArXiv preprint*, *arXiv*, 0601631.

Kauermann, G. (1996). On a dualization of graphical Gaussian models. *Scandinavian Journal of Statistics, 23(1)*, 105-116.

## Appendix C: Conditions to find decomposable graphs

The following proposition establishes which approaches will induce decomposable graphs. Hereinafter, the "functional blocks" mentioned in approach the approach considering the use of gene annotation will be referred to as blocks. In this approach, when effects of markers in different blocks are not allowed to be correlated, the corresponding strategy will be referred to as approach F1. On the other hand, when the effects of subsets or markers in different blocks are assumed to be correlated, the corresponding strategy will be referred to as approach F2.

If a block contains a subset of markers with effects correlated with the effects of a subset of markers in another block, these blocks are said to be linked. Let $B$ be the total number of blocks and $\mathcal{L}$ be the set of pairs of linked blocks. Let $\Psi$ be the set of blocks linked with at least two other blocks, $\forall\, l \in \Psi$ let $\Gamma_l$ be the set of blocks linked to block $l$ and $\forall\, a \in \Gamma_l$, let $C_{l_a}$ be the subset of markers in block $l$ whose effects are correlated with effects of a subset of markers in block $a$, $1 \leq a \leq B, a \neq l$.

*Proposition 1*

The graphs induced under approaches considering correlation of groups of nearby markers and approach F1, are decomposable. In addition the graph induced under the approach F2 is decomposable if there exists an ordering of markers $\sigma'$ that along with the edge set satisfy the following conditions.

**Condition 1.1** For all possible triplets of linked blocks $\{l, l', l''\}$ such that $C_{l_{l'}} \neq C_{l_{l''}}$, $C_{l'_l} \neq C_{l'_{l''}}$, $C_{l''_l} \neq C_{l''_{l'}}$, and the sets $I_l := C_{l_{l'}} \cap C_{l_{l''}}$, $I_{l'} := C_{l'_l} \cap C_{l'_{l''}}$ and $I_{l''} := C_{l''_l} \cap C_{l''_{l'}}$, are all non-empty, the following never happens: $\sigma'(i) > \sigma'(j) > \sigma'(k)$, $i \in C_{l_{l''}} \cap I_l^c, j \in C_{l'_l}$ or $i \in C_{l_{l''}}, j \in C_{l'_l} \cap I_{l'}^c$, and $k \in I_{l''}$; if there are triplets of linked blocks $\{l, l', l''\}$ such that exactly one of the three sets $\{I_l, I_{l'}, I_{l''}\}$, say $I_l$ is empty, then: $\min\{\sigma'(k), \sigma'(i), \sigma'(j)\} = \sigma'(k), \forall\, k \in C_{l_{l'}} \cup C_{l_{l''}}, \forall\, j \in I_{l'}\, \forall\, i \in I_{l''}$ and if exactly two of these sets, say $\{I_l, I_{l'}\}$ are empty, then for either $l$ or $l'$, say $l$, $\sigma'(k) < \sigma'(i)\, \forall k \in C_{l_{l'}} \cup C_{l_{l''}}, \forall i \in I_{l''}$. Superindex $C$ indicates the complement with respect to the index set of the corresponding block.

**Condition 1.2** For every possible triplet of blocks $\{l, l', l''\}$ the following does not happen: $\sigma'(k) < \sigma'(j) < \sigma'(i)$, $k \in I_l, j \in C_{l'_l}, i \in C_{l''_l}, C_{l'_{l''}} = \emptyset$.

**Condition 1.3** For every duplet of linked blocks $\{l, l'\}$ the following does not hold: $\exists\, i \in l, \{j, k\} \in l'$ such that $\sigma'(i) > \sigma'(j) > \sigma'(k), i \in C_{l_{l'}}, j \in C_{l'_l}^C, k \in C_{l'_l}$.

**Condition 1.4** For each pair of linked blocks $(l, l')$, $C_{l_{l'}} \times C_{l'_l} \in E_\sigma$, that is, the effect of each marker in $C_{l_{l'}}$ is correlated with the effects of all marker in $C_{l'_l}$.

Moreover, conditions 1.1, 1.2 and 1.3 are necessary whereas condition 1.4 is not.

This proposition involves all possible orderings of markers. However, if markers are ordered in such a way that markers in the same block are given consecutive indices, the number of possible orderings is reduced. Thus, in order to provide a simpler way to order markers, the following proposition only requires the existence of an ordering of the blocks and a structure on the edges set satisfying certain conditions that permit to find a perfect elimination ordering of markers.

*Proposition 2*

If there exists an ordering $\rho$ of the blocks which coupled with the structure of the edges set satisfy condition 1.4 plus the following conditions:

**Condition 2.1** $C_{l_a} = \cdots = C_{l_m} := C_l \ \forall \ l \in \Psi$

**Condition 2.2** For every possible triplet of blocks $\{l, l', l''\}$ the following does not happen: $(l, l'), (l, l'') \in L, (l', l'') \notin L, \rho(l) < \rho(l') < \rho(l'')$.

Then the following ordering strategy (denoted by $\sigma$) of marker loci is a perfect elimination ordering: once blocks have been ordered according to $\rho$, markers are ordered in such a way that the smaller the index of a block the smaller the indices of the markers pertaining to that block. The ordering inside each block is done as follows: markers in $C_l$ are given the largest indices in block $l$. In addition, under this ordering strategy, condition 2.2 is also necessary for $\sigma$ to be a perfect elimination ordering whereas condition 2.1 is not.

*Corollary to Proposition 2*

Consider the "super graph" formed by regarding the blocks as super nodes and $\mathcal{L}$ as a "super vertices set". Then, under conditions 2.1 and 1.4, if the "super vertices set" admits a perfect elimination ordering, the ordering defined in proposition 2 corresponds to a perfect elimination scheme.

**Proofs of propositions 1 and 2, and corollary to proposition 2**

**Proof of proposition 1**

Approach 1 induces either a banded or a differentially banded matrix and these structures are known to correspond to decomposable graphs. Similarly, 3.a induces block diagonal matrices which also correspond to decomposable graphs. The only non-trivial part is to prove the statements made about conditions for 3.b to induce a decomposable graph. To prove it, we show that it is possible to find a perfect elimination ordering, and we use the fact that the existence of a perfect elimination ordering characterizes decomposable graphs.

*Proof of sufficiency*

Hereinafter, consider arbitrary indices $\{i, j, k\}$ such that according to $\sigma'$ $i > j > k$.

**Case 1)** The triplet $\{i, j, k\}$ pertains to the same block. Because each block corresponds to a complete graph, it follows that in this case it cannot be that

$$i > j > k, (i,j) \notin E_\sigma, (i,k), (j,k) \in E_\sigma \quad (A.1)$$

**Case 2)** $i \in l, \{j, k\} \in l', (l, l') \in \mathcal{L}$. If $(i, j) \notin E_\sigma$, by condition 1.4 it means that one of the following mutually exclusive events must occur:

- $i \notin C_{l_{l'}}$ and $j \in C_{l'_l}$
- $i \in C_{l_{l'}}$ and $j \notin C_{l'_l}$
- $i \notin C_{l_{l'}}$ and $j \notin C_{l'_l}$.

It follows that:
$$i \notin C_{l_{l'}} \Rightarrow (i,k) \notin E_\sigma$$
this suffices to show that condition $A.1$ cannot be attained under the first and third scenarios. On the other hand:
$$i \in C_{l_{l'}} \text{ and } j \notin C_{l'_l} \Rightarrow k \notin C_{l'_l} \text{ ($\because$ condition 1.3)}$$
$$\Rightarrow (i,k) \notin E_\sigma,$$
which is enough to show that condition $A.1$ does not hold under the second scenario.

**Case 3)** $(i,j) \in l, k \in l'$. In this case $(i,j) \in E_\sigma$ because all blocks are complete; therefore, condition $A.1$ cannot be met.

**Case 4)** $i \in l, j \in l', k \in l''$. If $k \notin C_{l''_l} \cup C_{l''_{l'}}$, it is clear that $(j,k), (i,k) \notin E_\sigma \ \forall i \in B_l, \forall j \in B_{l'}$. If each one of the three blocks is linked with the other two, that is, $(l,l'), (l',l''), (l,l'') \in \mathcal{L}$, condition 1.1 implies that if all three intersection sets are non-empty and $(i,j) \notin E_\sigma$ then either $(i,k) \notin E_\sigma$ or $(j,k) \notin E_\sigma$ which prevents the occurrence of $A.1$. If one or two of the three intersection subsets is empty, by condition 1.1 it follows that $I_{l''} = \emptyset$ then:
$$k \in C_{l''_l} \Rightarrow (j,k) \notin E_\sigma$$
$$k \in C_{l''_{l'}} \Rightarrow (i,k) \notin E_\sigma,$$
Notice that this holds disregarding the location of indices $j$ and $i$ within their corresponding blocks; therefore, condition $A.1$ cannot be attained.

On the other hand, if one block is linked to the other two, and these are the only existent links, condition 1.2 guarantees that $A.1$ does not hold.

Until here, only linked blocks were considered. Notice that if at least one index of the triplet $\{i,j,k\}$ corresponds to a marker in an isolated (i.e. not linked) block, then $A.1$ does not happen.

In conclusion, under the four conditions of result 1 it is possible to find a perfect elimination ordering and therefore the graph induced under approach 3.b is decomposable.

Now we proceed to prove necessity of conditions 1,2 and 3. To this end, it is proven that under 3.b, if these conditions do not hold, then the induced graph is not decomposable.

*Proof of necessity of conditions 1.1, 1.2 and 1.3*

Hereinafter when the words "there is always" precede a condition, it is meant that under any ordering of markers, the graph always satisfies the condition.

Suppose that 1.1 does not hold. It means that:

- There is always at least one triplet of linked blocks $\{l, l', l''\}$, $1 \leq l, l', l'' \leq |\Psi|$, linked in the following way: $C_{l_{l'}} \neq C_{l_{l''}}$, $C_{l'_l} \neq C_{l'_{l''}}$, $C_{l''_l} \neq C_{l''_{l'}}$, the three pairwise intersections $I_l$, $I_{l'}$ and $I_{l''}$ are non-empty, and there exist at least one triplet $\{i,j,k\}$ such that $i > j > k$ and $i \in C_{l_{l''}} \cap I_l^c, j \in C_{l'_l}$ or $i \in C_{l_{l''}}, j \in C_{l'_l} \cap I_{l'}^c$, and $k \in I_{l''}, i > j > k,$

or

- There is always at least one triplet of linked blocks $\{l, l', l''\}$ and a triplet of indices $\{i,j,k\}$ such that $i > j > k$, exactly one or two of the three sets $\{I_l, I_{l'}, I_{l''}\}$ are empty and $k$ pertains to a non-empty intersection set.

It is easy to notice that in any of the two events described in the first case, condition A.1 is satisfied. On the other hand, in the second case, if $k$ pertains to the (non-empty) intersection set of a block, say $I_l$, then we can always pick $i \in C_{l''_l}$ and $j \in C_{l'_l}$ such that $(i,j) \notin E_\sigma$, then, because $k \in I_l$ it follows that $(i,k), (j,k) \in E_\sigma$, that is, $A.1$ holds.

Therefore, if 1.1 does not hold, a perfect elimination ordering does not exist and the graph is not decomposable. Similarly, if there is always at least one triplet of blocks $\{l, l', l''\}$ satisfying the converse of condition 1.2 then $A.1$ always holds, that is, a perfect elimination ordering does not exist and consequently the graph is not decomposable. Finally, if 1.3 does not hold then always exists at least a linked pair $\{l, l'\}$ such that $\exists\, i \in C_{l_{l'}}, j \in C^C_{l'_l} \cap l, k \in C_{l'_l}, i > j > k$ which immediately implies $A.1$.

*Proof of non-necessity of condition 1.4*

We refute the statement that condition 1.4 is necessary by constructing a counter-example. Suppose that a graph is comprised of two blocks. Let $L_\sigma$ be the subset of $E_\sigma$ containing edges comprised of one node pertaining to block 1 and one node from block 2. Also assume that block 1 contains nodes $\{a, b, c, d\}$, $C_{1_2} := \{b, c, d\}, C_{2_1} := \{w, z\}$ and $\mathcal{L} := \{(d, w), (c, w), (b, z), (c, z), (d, z)\}$. Consider any ordering satisfying the following: $\sigma(a) = 1, \sigma(b) = 2, \sigma(c) = 3, \sigma(d) = 4$. Consider $i > j > k, i \in C_{2_1}$, if $i = \sigma(w)$ then $(i,j) \notin E_\sigma \Rightarrow (i,k) \notin E_\sigma$, if $i = \sigma(z)$ then $(i,j) \notin E_\sigma \Rightarrow j = 1 = \sigma(a) \therefore \nexists\, k < j$ and consequently $A.1$ does not hold. We have found a family of perfect elimination orderings and its existence implies that the graph is decomposable even though condition 4 does not hold. It proves that condition 3 is not a necessary condition.

∎

**Proof of proposition 2**

*Proof of sufficiency*

The general approach is very similar to that used to prove sufficiency of conditions stated in result 1. Let $\Lambda$ be the set of blocks linked to at least one block. Assume that markers have been ordered as stated in result 2, this ordering is denoted by $\sigma$. Let $B^\sigma_l$ be the set of indices corresponding to markers in block $l$ under $\sigma$. Let $\{i, j, k\}$ be three arbitrary indices such that according to $\sigma$, $i > j > k$, and $C_l \in B^\sigma_l$, $l \in \Lambda$ be the subset of indices corresponding to markers in block $l$ whose effects are partially correlated with the effects of a subset of markers in another block $1 \leq l \leq |\Lambda|$. In the following, it is proven that under conditions 1.4, 2.1 and 2.2 this within-block ordering strategy yields a perfect elimination ordering. Four cases depending on the position of indices $i, j, k$ are considered.

**Case 1**) The triplet $\{i, j, k\}$ pertains to the same block. Because each block corresponds to a complete graph, it follows that in this case condition $A.1$ is not satisfied.

**Case 2**) $i \in l, \{j, k\} \in l', \{l, l'\} \in \Lambda, \rho(l) > \rho(l')$. If $(i,j) \notin E_\sigma$, by condition 1.4, it means that one of the following mutually exclusive events must occur:

- $i \notin C_l$ and $j \in C_{l'}$
- $i \notin C_l$ and $j \notin C_{l'}$
- $i \in C_l$ and $j \notin C_{l'}$

It follows that:
$$i \notin C_l \Rightarrow (i,k) \notin E_\sigma,$$

this suffices to show that condition $A.1$ cannot be attained under the first and second events. On the other hand:
$$i \in C_l \text{ and } j \notin C_{l'} \Rightarrow k \notin C_{l'} \ (\because k < j \text{ under } \sigma)$$
$$\Rightarrow (i, k) \notin E_\sigma,$$
which is enough to show that condition $A.1$ does not hold under the second scenario.

**Case 3)** $\{i, j\} \in l, k \in l', l > l'$. In this case $(i, j) \in E_\sigma$ because all blocks are complete; therefore, condition $A.1$ cannot be met.

**Case 4)** $i \in l, j \in l', k \in l'', \rho(l) > \rho(l') > \rho(l'')$. If $k \notin C_{l''}$ it is clear that $(j, k), (i, k) \notin E_\sigma \ \forall \ i \in B_l \ \forall \ j \in B_{l'}$ disregarding the position of indices $i$ and $j$ within their corresponding blocks and the linkage relationship among the three blocks, thus $A.1$ does not hold and, therefore; in the following, only the case $k \in C_{l''}$ is considered. If each one of the three blocks is linked with the other two, then:
$$(i, j) \notin E_\sigma \Rightarrow i \notin C_l \text{ or } j \notin C_{l'}$$
$$\Rightarrow (i, k) \notin E_\sigma \text{ or } (j, k) \notin E_\sigma \ (\because \text{ condition 2.1}).$$

Similarly, if one block is linked to the other two, and these are the only existent links, condition 2.2 guarantees that $A.1$ does not hold.

Consequently, under the two conditions of result 2 and condition 1.4, ordering $\sigma$ is a perfect elimination ordering. Thus, the graph induced under approach 3.b with and edges set as defined in result 2 is decomposable.

*Proof of necessity of condition 2.2*

To prove necessity of condition 2.2, we show that if it is not satisfied, then $\sigma$ is not a perfect elimination ordering. If there is at least one triplet of blocks $\{l, l', l''\}$ such that $(l, l'), (l, l'') \in \mathcal{L}, (l', l'') \notin \mathcal{L}, \rho(l) < \rho(l') < \rho(l'')$ then by picking $i \in C_{l''}^l, j \in C_{l'}^l$ and $k \in C_{l'}$ it follows that $i > j > k$, and $(i, j) \notin E_\sigma, (i, k), (j, k) \in E_\sigma$, that is, condition A.1 holds and as a consequence, $\sigma$ is not a perfect elimination ordering, which implies necessity of condition 2.2.

*Proof of non-necessity of condition 2.1*

By constructing a counter example, we refute the necessity of condition 2.1. Suppose that $G$ contains three blocks such that $\rho(B_1) = 1, \rho(B_2) = 2, \rho(B_3) = 3$, $B_1$ is only linked to $B_2$ and $B_2$ is only linked to $B_3$, $C_{2_1} \neq C_{2_3}$ and markers have been ordered according to $\sigma$. It is easy to see that in this graph, for any triplet $\{i, j, k\}$ such that according to $\sigma, i > j > k$ if $(i, j) \notin E_\sigma$ then $(i, k) \notin E_\sigma$ which implies that condition A.1 is never reached and consequently, $\sigma$ is a perfect elimination ordering despite of the fact that condition 2.1 does not hold. Consequently, 2.1 is not a necessary condition.

∎

**Proof of corollary to proposition 2**

Let $G^s$ be the "super graph" defined in the corollary. It is assumed that conditions 2.1 and 1.4 hold. Consequently, we only need to check that condition 2.2 is satisfied. It follows immediately by noticing that condition 2.2 is nothing but the definition of a perfect elimination ordering of the blocks, i.e., a perfect elimination ordering of the "super vertices" of $G^s$.

∎